\documentclass{appolb}
\usepackage[colorlinks,linkcolor=blue,citecolor=red]{hyperref}
\usepackage{graphicx}

\begin{document}
\title{Proton number cumulants and correlation functions
from hydrodynamics and the QCD phase diagram%
\thanks{Presented at Quark Matter 2022}%
}

\author{Volodymyr Vovchenko\\[-3mm]
\address{Institute for Nuclear Theory, University of Washington, Box 351550, Seattle, WA 98195, USA\\
Nuclear Science Division, Lawrence Berkeley National Laboratory, 1 Cyclotron Road,  Berkeley, CA 94720, USA}
\\[4mm]
{Volker Koch\\[-3mm]
\address{Nuclear Science Division, Lawrence Berkeley National Laboratory, 1 Cyclotron Road,  Berkeley, CA 94720, USA}
}
\\[4mm]
Chun Shen\\[-3mm]
\address{Department of Physics and Astronomy, Wayne State University, Detroit, Michigan 48201, USA\\
RIKEN BNL Research Center, Brookhaven National Laboratory, Upton, NY 11973, USA}
}

\maketitle
\begin{abstract}
We analyze the behavior of (net-)proton number cumulants in central collisions of heavy ions across a broad collision energy range by utilizing hydrodynamic simulations. 
The calculations incorporate essential non-critical contributions to proton fluctuations such as repulsive baryonic core and exact baryon number conservation.
The experimental data are consistent with non-critical physics at collision energies $\sqrt{s_{\rm NN}} \geq 20$~GeV.
The data from STAR and HADES Collaborations at lower collision energies indicate an excess of (multi-)proton correlations over the non-critical reference.
This observation is discussed in the context of different mechanisms, including the possibility of a critical point in the baryon-rich region of the QCD phase diagram.
\end{abstract}
  
\section{Introduction}
Determination of the phase structure of QCD matter is one of the key questions tackled by relativistic heavy-ion collisions at various energies~\cite{Bzdak:2019pkr}.
Of particular relevance is the location~(and even the existence) of the QCD critical point at finite baryon density.
Proton number fluctuations are considered to be particularly sensitive probes of the QCD critical point~\cite{Hatta:2003wn}, especially high-order non-Gaussian measures~\cite{Stephanov:2008qz,Stephanov:2011pb}.
Experimental measurements have now been performed in a broad collision energy range by multiple experimental collaborations, including ALICE~\cite{ALICE:2019nbs}, STAR~\cite{STAR:2020tga,STAR:2021iop,STAR:2021fge}, and HADES~\cite{HADES:2020wpc}.
Measurements from the RHIC-BES program in particular indicate a possible non-monotonic collision energy dependence of net-proton kurtosis $\kappa \sigma^2$~\cite{STAR:2020tga} -- a potential signature of the QCD critical point~\cite{Stephanov:1999zu}.
However, the experimental uncertainties are too large to make a definitive conclusion on $\kappa \sigma^2$, this will be improved with the upcoming data from the BES-II program.
One can, however, ask a question if something can be learned instead from the more accurate measurements of the second- and third-order proton cumulants that are available.

Quantitative comparisons between experimental measurements and theoretical expectations of event-by-event fluctuations are challenging due to the many caveats involved~\cite{Vovchenko:2021gas}, and thus require extensive dynamical modeling of heavy-ion collisions.
In the context of the search for the QCD critical point, one can consider different strategies.
Ideally, one would incorporate critical fluctuations into the hydrodynamic framework for heavy-ion collisions, and make testable predictions based on the location of the critical point. Such a framework is currently under development~\cite{Bluhm:2020mpc,An:2021wof}.
Alternatively, one could use a microscopic approach such as, for example, molecular dynamics with a critical point. Some promising recent developments in that regard can be found in Refs.~\cite{Sorensen:2020ygf,Kuznietsov:2022pcn} which are, however, not yet ready for applications to expanding systems created in heavy-ion collisions.
Finally, one could study deviations of the experimental data from precision calculations of non-critical contributions to proton number cumulants. This approach was adopted in~\cite{Vovchenko:2021kxx} and is the focus of the present work.

\section{Proton number cumulants from hydrodynamics}
Hydrodynamics provides a realistic background over which non-critical contributions to proton number cumulants can be calculated. 
These contributions include (i) the exact conservation of baryon number~\cite{Bzdak:2012an} and (ii) the repulsive core in the baryon-baryon interaction~\cite{Vovchenko:2017xad}.
These effects are incorporated at the Cooper-Frye particlization stage, either through analytic approximations~\cite{Vovchenko:2021kxx} or by using Monte Carlo sampling~\cite{Vovchenko:2020kwg}.

\subsection{LHC}

ALICE Collaboration has measured the normalized variance $\kappa_2[p-\bar{p}]/\langle p+\bar{p} \rangle$ of the net-proton distribution in 2.76~TeV Pb-Pb collisions~\cite{ALICE:2019nbs}.
The data show a subtle suppression of this quantity with respect to the baseline of unity, which increases with the pseudorapidity acceptance.
Neglecting the dynamics of the hadronic phase, this suppression is consistent with long-range, essentially global, conservation of the baryon number, while the additional effect of excluded volume cannot be distinguished within the presently available experimental uncertainties~\cite{Vovchenko:2020kwg}.

However, the agreement with the data breaks down when baryon annihilation in the hadronic phase modeled by UrQMD is incorporated simultaneously with global baryon conservation~\cite{Savchuk:2021aog}. 
In this case, the agreement is recovered if a more local (in the rapidity space) conservation of the baryon number is imposed instead. 
Both the range of exact baryon conservation and the role baryon annihilation can be constrained experimentally by a precise combined measurement of  $\kappa_2[p-\bar{p}]/\langle p+\bar{p} \rangle$ and  $\kappa_2[p+\bar{p}]/\langle p+\bar{p} \rangle$.
Additional constraints 
could be obtained from fluctuation measurements involving light nuclei~\cite{ALICE:2022amd}, as well as data-driven approaches~\cite{RustamovQM2022}.

\subsection{RHIC-BES}

\begin{figure}[!t]
\centering
\includegraphics[width=0.49\textwidth]{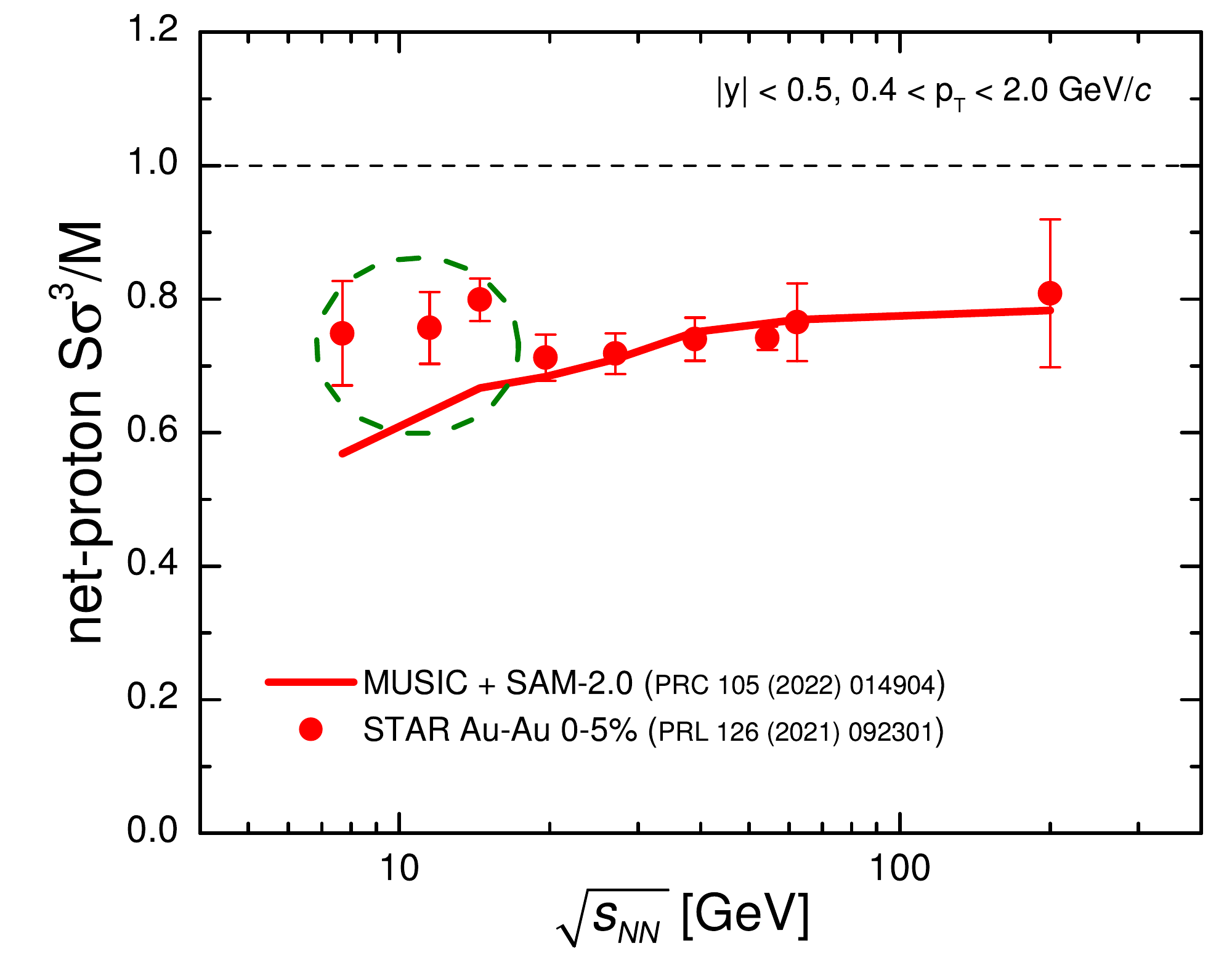}
\includegraphics[width=0.49\textwidth]{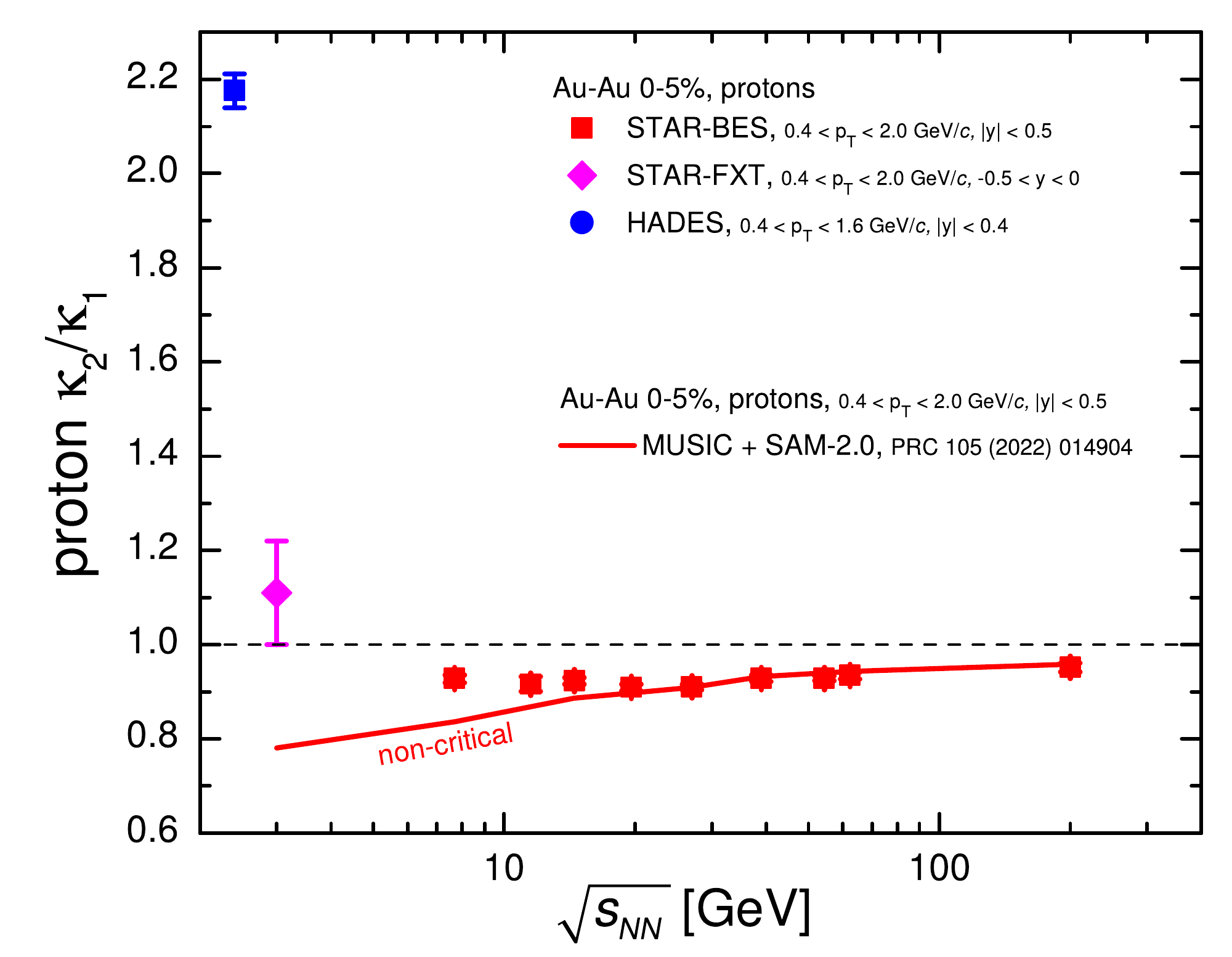}
\caption[]{
Beam energy dependence net-proton cumulant ratio $\kappa_3/\kappa_1$~(left panel) and proton scaled variance $\kappa_2/\kappa_1$~(right panel) as calculated based on hydrodynamics framework MUSIC and correction for baryon conservation via a method SAM-2.0.
The symbols correspond to experimental data from STAR~\cite{STAR:2020tga,STAR:2021iop} and HADES~\cite{HADES:2020wpc} Collaborations.
Adapted from Ref.~\cite{Vovchenko:2021kxx}.
}\label{fig:RHIC}
\end{figure}

At RHIC-BES energies we calculate the (net-)proton cumulants analytically~\cite{Vovchenko:2021kxx}, based on MUSIC simulations of 0-5\% central Au-Au collisions~\cite{Shen:2020jwv}. 
As before, the calculations incorporate the effect of baryon repulsion by means of excluded volume, as well as correction for global baryon conservation by a method called SAM-2.0~\cite{Vovchenko:2021yen}.
The results are shown in Fig.~\ref{fig:RHIC}.
Experimental data on the net-proton cumulant ratio $\kappa_3/\kappa_1$~\cite{STAR:2020tga} show suppression of this ratio relative to the uncorrelated proton production baseline of unity at all collision energies.
At $\sqrt{s_{\rm NN}} \geq 20$~GeV the data are quantitatively described by our calculation when both baryon repulsion and baryon conservation are incorporated simultaneously.
The data at lower collision energies, however, indicate an excess of $\kappa_3/\kappa_1$ over our non-critical reference, indicated by the green circle in Fig.~\ref{fig:RHIC}.

It can be even more instructive to look at the second order cumulants, namely the scaled variance of proton number distribution which has been measured even more accurately.
As in the case of $\kappa_3/\kappa_1$, our model calculations are in good agreement with the STAR data~\cite{STAR:2021iop} at $\sqrt{s_{\rm NN}} \geq 20$~GeV.
However, the excess of two-proton correlations at lower collision energies is evident, where the model fails to capture both the magnitude and the slope of the collision energy dependence. This is especially true as one goes to even lower collision energies like $\sqrt{s_{\rm NN}} = 3$~GeV from STAR-FXT~\cite{STAR:2021fge} or $\sqrt{s_{\rm NN}} = 2.4$~GeV from SIS-HADES~\cite{HADES:2020wpc}.
The non-critical calculations, extended down to $\sqrt{s_{\rm NN}} = 3$~GeV by means of the blast-wave model, show no change in the trend for $\kappa_2/\kappa_1$ whereas the experimental data, in particular at HADES, show a dramatic enhancement of the scaled variance of proton number.

Our calculations have neglected the possible effect of volume fluctuations~\cite{Gorenstein:2011vq,Skokov:2012ds}. The description of the RHIC-BES data at $\sqrt{s_{\rm NN}} \leq 20$~GeV could be improved by adding volume fluctuations via an additional parameters, but this would then spoil the agreement at higher collision energies. No improvement can be obtained by incorporating the additional effect of exact conservation of electric charge~\cite{Vovchenko:2021kxx}.

It should also be noted the experimental data from RHIC-BES indicate sizable negative two-particle correlations among the antiprotons.
This behavior is reproduced qualitatively by baryon conservation and excluded volume effects, but the magnitude of the correlations is notably underestimated at most of the collision energies.

\subsection{SIS-HADES}

\begin{figure}[!t]
\centering
\includegraphics[width=0.49\textwidth]{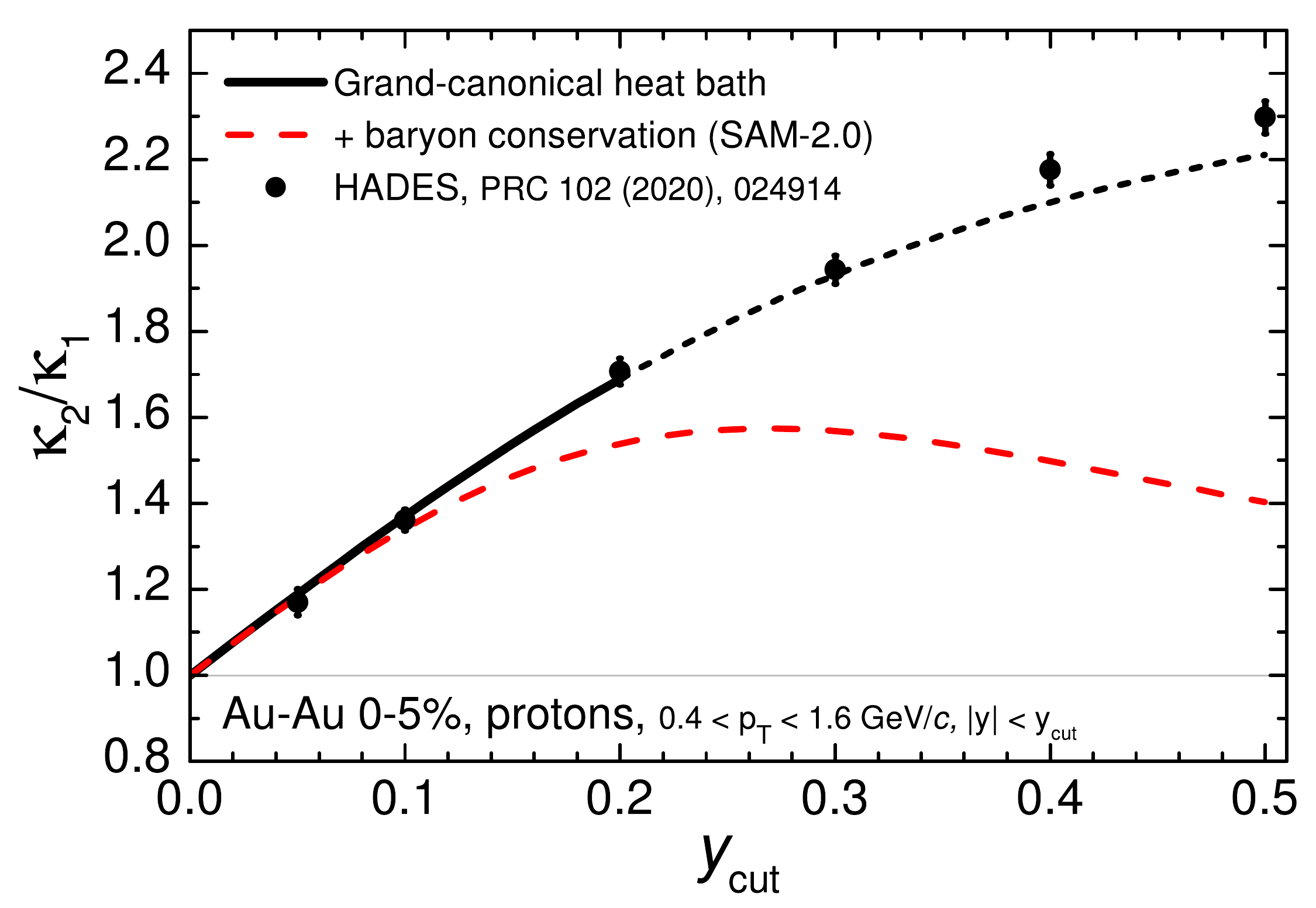}
\includegraphics[width=0.49\textwidth]{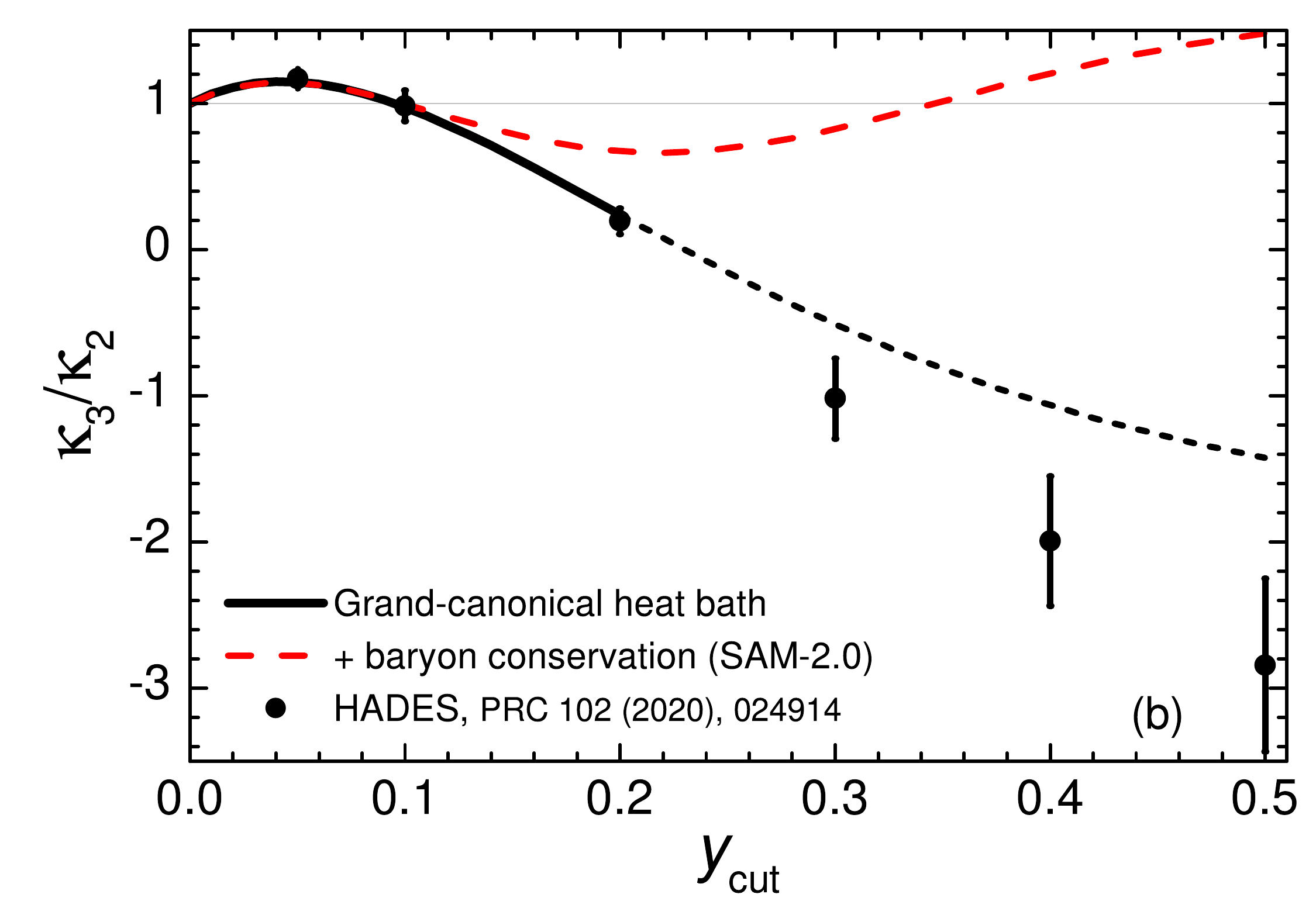}
\caption[]{
Rapidity cut dependence of proton number cumulants $\kappa_2/\kappa_1$~(left panel) and $\kappa_3/\kappa_2$~(right panel) in 0-5\% central Au-Au collisions at $\sqrt{s_{\rm NN}} = 2.4$~GeV calculated assuming thermal emission from a Siemens-Rasmussen like fireball with and without the effect of exact baryon conservation.
Adapted from Ref.~\cite{Vovchenko:2022szk}.
}\label{fig:HADES}
\end{figure}

The strong increase in proton $\kappa_2/\kappa_1$ at $\sqrt{s_{\rm NN}} = 2.4$~GeV warrants a closer look at the HADES data. In~\cite{Vovchenko:2022szk} these data were analyzed as a function of rapidity cut $y_{\rm cut}$ within the framework of a Siemens-Rasmussen like fireball model, with chemical and kinetic freeze-out parameters based on Refs.~\cite{Harabasz:2020sei,Motornenko:2021nds}.
The modeling also incorporates the fact that a significant fraction of protons are bound into light nuclei and do not contribute to the measured cumulants of proton number.
The analysis found that the HADES data~\cite{HADES:2020wpc} on the rapidity acceptance of the proton cumulant ratios up to the fourth order can be described by assuming a thermal emission of nucleons from a grand-canonical heat bath~(Fig.~\ref{fig:HADES}), provided that the corresponding baryon number susceptibilities of QCD matter characterizing the emitting source are highly non-Gaussian and exhibit the following hierarchy:
$\chi_4^B \gg -\chi_3^B \gg \chi_2^B \gg \chi_1^B$.
Naively, this observation could point to a presence of the QCD critical point close to the HADES chemical freeze-out at $T \sim 70$~MeV and $\mu_B \sim 850-900$~MeV, given that the critical point is one such potential source of non-Gaussian fluctuations~\cite{Stephanov:2008qz}.

However, the grand-canonical picture can only be applicable when the effect of exact baryon conservation is small, for instance by considering small rapidity acceptance cuts $y_{\rm cut} \leq 0.2$.
When the exact baryon conservation is incorporated into the calculations, here done by means of the SAM-2.0 method~\cite{Vovchenko:2021yen}, the description of the data becomes challenging for $y_{\rm cut} \geq 0.2$~(dashed red lines in Fig.~\ref{fig:HADES}).
This indicates that more theoretical and experimental effort is required to reach a firm conclusion.

\section{Summary}

The available experimental data on (net-)proton cumulants in central collisions of heavy ions are shown to be consistent at $\sqrt{s_{\rm NN}} \geq 20$~GeV with non-critical physics such as baryon number conservation and short-range repulsion incorporated on top of standard hydrodynamical description. The data from STAR and HADES Collaborations at lower collision energies, on the other hand, indicate an excess of (multi-)proton correlations over the non-critical reference. The critical point in the baryon-rich region of the QCD phase diagram has been discussed here as a possible mechanism behind the excess, while pointing out that other possible explanations such as volume fluctuations have to be carefully analyzed as well. It is also noted that the large non-Gaussian fluctuations observed in the HADES experiment are challenging to describe in the context of baryon number conservation irrespective of the underlying model.


\begin{thebibliography}{}


\bibitem{Bzdak:2019pkr}
A.~Bzdak, S.~Esumi, V.~Koch, J.~Liao, M.~Stephanov and N.~Xu,
Phys. Rept. \textbf{853} (2020), 1-87.

\bibitem{Hatta:2003wn}
Y.~Hatta and M.~A.~Stephanov,
Phys. Rev. Lett. \textbf{91} (2003), 102003.

\bibitem{Stephanov:2008qz}
M.~A.~Stephanov,
Phys. Rev. Lett. \textbf{102} (2009), 032301.

\bibitem{Stephanov:2011pb}
M.~A.~Stephanov,
Phys. Rev. Lett. \textbf{107} (2011), 052301.

\bibitem{ALICE:2019nbs}
S.~Acharya \textit{et al.} [ALICE],
Phys. Lett. B \textbf{807} (2020), 135564.

\bibitem{STAR:2020tga}
J.~Adam \textit{et al.} [STAR],
Phys. Rev. Lett. \textbf{126} (2021) 092301.

\bibitem{STAR:2021iop}
M.~Abdallah \textit{et al.} [STAR],
Phys. Rev. C \textbf{104} (2021) 024902.

\bibitem{STAR:2021fge}
M.~S.~Abdallah \textit{et al.} [STAR],
Phys. Rev. Lett. \textbf{128} (2022) 202303.

\bibitem{HADES:2020wpc}
J.~Adamczewski-Musch \textit{et al.} [HADES],
Phys. Rev. C \textbf{102} (2020) 024914.

\bibitem{Stephanov:1999zu}
M.~A.~Stephanov, K.~Rajagopal and E.~V.~Shuryak,
Phys. Rev. D \textbf{60} (1999), 114028.

\bibitem{Vovchenko:2021gas}
V.~Vovchenko,
[arXiv:2110.02446 [nucl-th]].

\bibitem{Bluhm:2020mpc}
M.~Bluhm, 
\textit{et al.}
Nucl. Phys. A \textbf{1003} (2020), 122016.

\bibitem{An:2021wof}
X.~An, 
\textit{et al.}
Nucl. Phys. A \textbf{1017} (2022), 122343.

\bibitem{Sorensen:2020ygf}
A.~Sorensen and V.~Koch,
Phys. Rev. C \textbf{104} (2021) 034904.

\bibitem{Kuznietsov:2022pcn}
V.~A.~Kuznietsov, O.~Savchuk, M.~I.~Gorenstein, V.~Koch and V.~Vovchenko,
Phys. Rev. C \textbf{105} (2022) 044903.

\bibitem{Vovchenko:2021kxx}
V.~Vovchenko, V.~Koch and C.~Shen,
Phys. Rev. C \textbf{105} (2022) 014904.

\bibitem{Bzdak:2012an}
A.~Bzdak, V.~Koch and V.~Skokov,
Phys. Rev. C \textbf{87} (2013) 014901.

\bibitem{Vovchenko:2017xad}
V.~Vovchenko, A.~Pasztor, Z.~Fodor, S.~D.~Katz and H.~Stoecker,
Phys. Lett. B \textbf{775} (2017), 71-78.

\bibitem{Vovchenko:2020kwg}
V.~Vovchenko and V.~Koch,
Phys. Rev. C \textbf{103} (2021) 044903.

\bibitem{Savchuk:2021aog}
O.~Savchuk, V.~Vovchenko, V.~Koch, J.~Steinheimer and H.~Stoecker,
Phys. Lett. B \textbf{827} (2022), 136983.

\bibitem{ALICE:2022amd}
ALICE Collaboration,
[arXiv:2204.10166 [nucl-ex]].

\bibitem{RustamovQM2022}
A.~Rustamov, \emph{these proceedings}.

\bibitem{Shen:2020jwv}
C.~Shen and S.~Alzhrani,
Phys. Rev. C \textbf{102} (2020) 014909.


\bibitem{Vovchenko:2021yen}
V.~Vovchenko,
Phys. Rev. C \textbf{105} (2022) 014903.

\bibitem{Gorenstein:2011vq}
M.~I.~Gorenstein and M.~Gazdzicki,
Phys. Rev. C \textbf{84} (2011), 014904.

\bibitem{Skokov:2012ds}
V.~Skokov, B.~Friman and K.~Redlich,
Phys. Rev. C \textbf{88} (2013), 034911.

\bibitem{Vovchenko:2022szk}
V.~Vovchenko and V.~Koch,
[arXiv:2204.00137 [hep-ph]].

\bibitem{Harabasz:2020sei}
S.~Harabasz, W.~Florkowski, T.~Galatyuk, M.~Gumberidze, R.~Ryblewski, P.~Salabura and J.~Stroth,
Phys. Rev. C \textbf{102} (2020) 054903.

\bibitem{Motornenko:2021nds}
A.~Motornenko, J.~Steinheimer, V.~Vovchenko, R.~Stock and H.~Stoecker,
Phys. Lett. B \textbf{822} (2021), 136703.


\end{thebibliography}
\end{document}